\documentclass[12pt]{article}
\usepackage{amssymb,amsmath,amscd}

\usepackage[dvips]{graphicx}
\usepackage{latexsym}

\newcommand{\Z}{{\mathbb Z}}
\newcommand{\R}{{\mathbb R}}

\begin{document}

\topmargin -2pt


\headheight 0pt

\begin{flushright}
{\tt KIAS-P08063}
\end{flushright}

\vspace{5mm}

\begin{center}
{\Large \bf Rotating Black Hole Entropy \\from Two Different Viewpoints} \\

\vspace{10mm}

{\sc Ee Chang-Young}${}^{ \dag,
 }$\footnote{cylee@sejong.ac.kr}, {\sc Daeho Lee}${}^{  \dag,  }$\footnote{dhlee@sju.ac.kr}, and {\sc Myungseok Yoon}${}^{  \ddag,  }$\footnote{younms@sogang.ac.kr}
\\

\vspace{1mm}

 ${}^{\dag}${\it Department of Physics, Sejong University, Seoul 143-747, Korea}\\

${}^{\ddag}${\it Center for Quantum Spacetime, Sogang
    University, Seoul 121-742, Korea}\\

\vspace{10mm}
{\bf ABSTRACT}
\end{center}
Using the brick-wall method, we study the entropy of
Kerr-Newman black hole from two different viewpoints,
a rest observer at infinity and zero angular momentum
observer near horizon.
We investigate this with scalar field in the canonical quantization approach.
An observer at infinity can take  one of the two possible frequency ranges;
one is with positive frequencies only and
the other is with the whole range including negative frequencies.
On the other hand, a zero  angular momentum
observer near horizon can take positive frequencies only.
For the observer at infinity the superradiant modes
appear in either choice of the frequency ranges and the two results coincide.
For the zero  angular momentum
observer superradiant modes do not appear due to absence of ergoregion.
The resulting entropies from the two viewpoints turn out to be the same.


\noindent

\vfill

\noindent

\thispagestyle{empty}

\newpage


%

\section{Introduction}
\label{intro}

Since Bekenstein suggested that a black hole has an intrinsic
entropy proportional to the surface area of its event horizon
\cite{bekenstein}, there have been many works to explain its statistical
origin \cite{zt}. Along this line, the ``brick-wall
model'' proposed by 't Hooft \cite{thooft} is to calculate the
entropy of a black hole by counting the degrees of freedom near its horizon.
By introducing a cutoff, the divergence due to the infinite blue shift near
horizon is removed \cite{mukohyama,su,dlm}. Note that a
global thermal equilibrium between the black hole and its surrounding
is assumed in this model. Therefore, the original brick-wall model
cannot be applied to a non-equilibrium system. However, the degrees of freedom
are mostly concentrated near the horizon, it is good enough to consider
only the narrow region near horizon
which is locally in thermal equilibrium with the black hole.
In this context, the ``thin-layer method" as an improved
brick-wall method has been also introduced \cite{hzk,zl,kpsy}. In
the thin-layer method, the local thermal
equilibrium is assumed. Assuming this kind of local thermal equilibrium
near horizon, one can
calculate the entropies of various black holes.

When one applies this method to the rotating black hole case,
one encounters the so-called superradiant modes.
The superradiant modes are special mode solutions of scalar fields
that satisfy the Klein-Gordon
equation in a given background spacetime of a rotating black hole,
and their appearance is due to the presence of the ergoregion
in which a particle cannot remain at rest as viewed from
an observer at infinity.
Counting these superradiant modes in the rotating case has caused
some confusion in the evaluation of the entropy.
Considering only non-superradiant modes, the
entropies of various rotating black holes were evaluated in
\cite{lk:pla,lk:prd}.
Because of the divergence of the free energy
from large azimuthal quantum number, they did not consider the
entropy contribution from the superradiant modes.
 In \cite{kkps,hkps}, an extra cutoff parameter was introduced in order to
 overcome the above divergence from the superradiant modes,
and it yielded incorrect answers.

In \cite{hk}, the superradiant modes were dealt
 with the correct quantization from the viewpoint of an observer at infinity
 in the rotating BTZ black hole case,
  and it was shown that the leading order divergence from the superradiant modes cancels
   the leading order divergence from the non-superradiant modes.
  Rather recently,  Kenmoku {\it et. al.} \cite{kins}
  evaluated the scalar field contribution to rotating black hole entropy
in an arbitrary $D$ dimensional spacetime.
Although their result contains the contribution from the superradiant modes,
in their calculation they did not separate the contributions from the superradiant
and the non-superradiant modes.
 In the rotating BTZ black hole case, their result coincides with the one obtained in \cite{hk}.
  In the four dimensional Kerr-Newman black hole case, besides the result of \cite{kins}
  so far there has been no correct calculation of the entropy from the superradiant part
  from the conventional viewpoint as in \cite{hk} in which the superradiant modes
  are considered to have positive frequencies.

In this paper, we reanalyze these results critically from a consistent setting,
and calculate the entropy of the Kerr-Newman black hole from the viewpoint
of a rest observer at infinity (ROI)
following the canonical quantization approach given in \cite{ford}
as it was done in \cite{hk}.
This result coincides with the result of \cite{kins}.
We then calculate the entropy of the Kerr-Newman black hole from the viewpoint
of a zero angular momentum observer (ZAMO) near horizon.
The result coincides with the one from the  ROI's viewpoint.

The organization of the paper is as follows.
In section 2, we calculate the entropy of a rotating black hole with a scalar field
in the canonical quantization approach from the ROI's viewpoint.
In section 3, we calculate the entropy of a rotating black hole from ZAMO's viewpoint.
In section 4, we conclude with discussion.


\section{Entropy from the viewpoint of ROI}

In this section we calculate the entropy of the
Kerr-Newman (KN) black hole from the viewpoint of ROI
 using the brick-wall method with a massless real scalar field.
The Kerr-Newman black hole solution is given by \cite{hkps,kins,kang}
\begin{eqnarray}
\label{metric}
ds^2= g_{tt} dt^2 +2g_{t\phi} dtd\phi
+g_{\phi\phi}d\phi^2 +g_{rr}dr^2
+g_{\theta\theta}d\theta^2,
\end{eqnarray}
where
\begin{eqnarray}
\label{metriccomp}
g_{tt} &=& -\frac{\Delta-a^2\sin^2\theta}{\Sigma},~~~~~~
g_{t\phi}= -\frac{a\sin^2\theta(r^2+a^2-\Delta)}{\Sigma},
\nonumber \\
g_{\phi\phi} &=&
\frac{(r^2+a^2)^2-\Delta a^2\sin^2\theta}{\Sigma}\sin^2\theta,
~g_{rr} = \frac{\Sigma}{\Delta},~g_{\theta\theta}=\Sigma,
\end{eqnarray}
and
\begin{eqnarray}
  \Sigma=r^2+a^2\cos^2\theta, ~\Delta=r^2-2Mr+a^2+Q^2.
\end{eqnarray}
Here $M$, $a$, $Q$ are  mass, angular momentum per unit mass,
and charge of the black hole, respectively.

The Kerr-Newman black hole has two coordinate singularities,
the outer and inner horizons, $r_{\pm}=M \pm \sqrt{M^2-a^2-Q^2}$ subject to
a condition $M^2 \ge a^2+Q^2$. The outer horizon
is defined as the event horizon. The Kerr-Newman metric has a stationary
limit surface which is the boundary of the ergoregion defined by
$r_{erg}=M+\sqrt{M^2-a^2\cos^2\theta-Q^2}$.
In the ergoregion a particle cannot remain at rest
as viewed from ROI.

 Now we consider a quantum gas of scalar particles
 confined in a box near the horizon.
 With the metric (\ref{metric}), the matter action for a massless real scalar field $\Phi$ is given by
\begin{eqnarray}
\label{action}
I_{\texttt{matt}}=\int d^{4}x \sqrt{-g}
\left(-\frac{1}{2}\partial_{\mu}\Phi\partial^{\mu}\Phi
\right).
\end{eqnarray}
The resulting equation of motion is given by
\begin{eqnarray}
\label{kle}
\frac{1}{\sqrt{-g}} \partial_{\mu}(\sqrt{-g} \partial^{\mu}\Phi)=0
\end{eqnarray}
with the boundary conditions
\begin{eqnarray}
\label{bc}
\Phi(x)=0~~\texttt{for}~~ r\leq r_{+}+\epsilon~\texttt{and}~~ r\geq L.
\end{eqnarray}
Here, $r_{+}$ is the outer horizon and  $r_{+}+\epsilon$ and $L$
are the boundaries of the spherical box assumed in the brick-wall
model. We assume that the quantum gas is in thermal equilibrium with
a black hole at temperature $T$ and constant angular velocity
$\Omega$ with respect to a ROI inside the spherical box.

In the WKB approximation, assuming the wave function to be $\Phi(x)=e^{-i\omega t+i m\phi+iK(r,\theta)}$,
the field equation (\ref{kle}) yields the following constraint condition
\begin{eqnarray}
\label{ccon}
\frac{p_{r}^2}{g_{rr}}+\frac{p_{\theta}^2}{g_{\theta\theta}}
=\frac{1}{-\Gamma}(
\omega^2 g_{\phi\phi}+2\omega m g_{t\phi}+m^2g_{tt}),
\end{eqnarray}
where $p_{r}=\partial K/\partial r$, $p_{\theta}=\partial K/\partial \theta$,
$-\Gamma \equiv g_{t\phi}^2-g_{tt}g_{\phi\phi}=\Delta \sin^2\theta \geq 0$ for $r \geq r_{+}$,
and $p_{\phi}=m$ is the azimuthal quantum number.
The constraint condition (\ref{ccon}) can be rewritten as
\begin{eqnarray}
\label{reccon}
\frac{p_{r}^2}{g_{rr}}+\frac{p_{\theta}^2}{g_{\theta\theta}}
=\frac{g_{\phi\phi}}{-\Gamma}
(\omega-m\Omega_{+})(\omega-m\Omega_{-}),
\end{eqnarray}
where
\begin{eqnarray}
\label{avop}
\Omega_{\pm}(r,\theta)=-\frac{g_{t\phi}}{g_{\phi\phi}}
\pm \sqrt{\left(\frac{g_{t\phi}}{g_{\phi\phi}}\right)^2 -\frac{g_{tt}}{g_{\phi\phi}}}
\end{eqnarray}
are the maximum and minimum angular velocities that a
particle can have. The limited range of the angular velocity is due to
the restriction that no particle can move faster
than light.

According to the semiclassical quantization rule, the number of modes
is given by \cite{thooft}
\begin{eqnarray}
\label{moden}
\pi n(\omega,m) \equiv
\int_{r_{+}+\epsilon}^{L} dr k(r; \omega,m).
\end{eqnarray}
%
%
In order to get the expression for $k(r; \omega,m)$
we first define the number of modes $n(\omega)$ with energy not exceeding $\omega$ as the sum over
the phase space divided by the unit quantum volume
$(2\pi)^3$ (with $\hbar=1$):
\begin{eqnarray}
\label{m1}
(2\pi)^3 n(\omega)= \int dr dp_r d\theta dp_{\theta}d\phi dp_{\phi}.
\end{eqnarray}
%
 After integrating
 over the momenta $p_{r}$ and $p_{\theta}$ and setting $p_{\phi}=m$,
 the number of modes $n(\omega)$ can be expressed as follows:
\begin{eqnarray}
\label{m2}
n(\omega)=\frac{1}{8\pi^2} \int dr d\theta d\phi dm \sqrt{g_{rr}g_{\theta\theta}}
\left(\frac{g_{\phi\phi}}{-\Gamma}\right)(\omega-m\Omega_{+})(\omega-m\Omega_{-}).
\end{eqnarray}
Note that in performing the above integration a condition
that the right hand side of Eq. \eqref{reccon} be positive should be satisfied.
This restriction comes from the way we evaluate the $p_\theta$ and  $p_r$ integral
by calculating the area of the $p_\theta$-$p_r$ ellipse satisfying the condition (8).
\begin{figure}[pt]
  \includegraphics[width=0.85\textwidth]{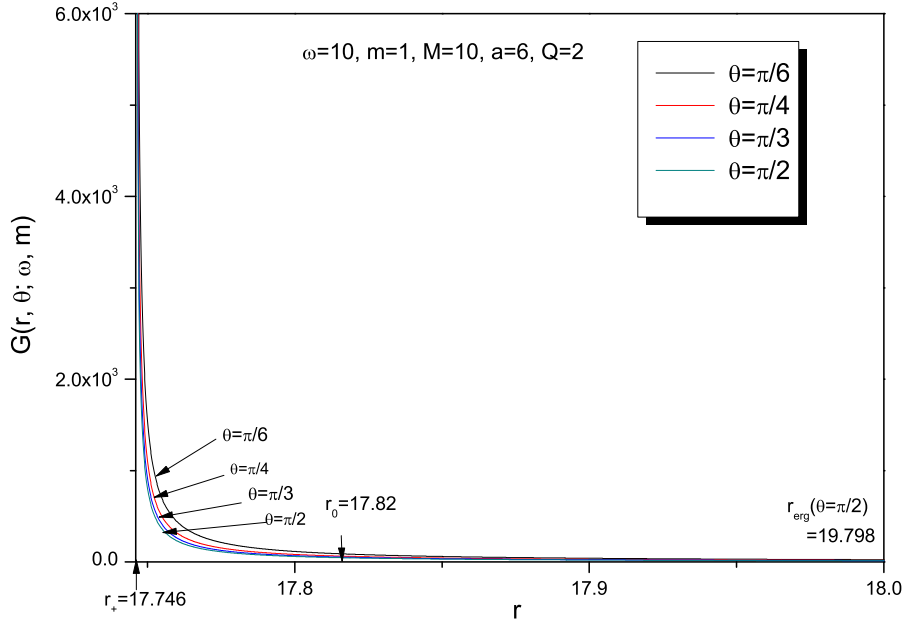}
  \caption{The number of modes per invariant volume is shown at various polar angles.
  It's not changed by $\theta \rightarrow \pi-\theta$.
  Here, $r_0$ is a position at which $\Omega_{0}$ differs by $1\% $ from
  $\Omega_{H}$ (See Fig.~\ref{fig:angv}). }
  \label{fig:dof}
\end{figure}
%
%
Therefore the radial wave number $k(r; \omega,m)$ defined in \eqref{moden} is given as follows
 with the restriction that the right hand side of Eq. \eqref{reccon}
be positive:
\begin{eqnarray}
\label{nk}
k(r; \omega,m)=\int \int d\theta d\phi~N^{-2} (\omega-m\Omega_{+})(\omega-m\Omega_{-}),
\end{eqnarray}
where
\[ N^{-2}= (-g^{tt})(g_{rr}g_{\theta\theta})^{1/2}/8\pi ~~{\rm with}~~ g^{tt}=g_{\phi\phi}/\Gamma.\]
%

%
Since $k(r;\omega,m) \propto N^{-2}$ and $N \rightarrow 0$ near the horizon,
the dominant contribution
in Eq. (\ref{moden}) comes from integrating near horizon region
as the inner boundary of the brick-wall approaches to it.
%
We can see this by
evaluating the number of modes per invariant volume $G(r,\theta;\omega,m)$ for given
$\omega$ and $m$, which is given by the following relation:
\begin{eqnarray}
\label{dof}
n(\omega,m)=\int d\mu_{\texttt{inv}}~G(r,\theta;\omega,m),
\end{eqnarray}
where
\[
G(r,\theta;\omega,m)
=\frac{1}{8\pi^2} \sqrt{g_{\phi\phi}}(\omega-m\Omega_{+})(\omega-m\Omega_{-})/(-\Gamma)  ~~ {\rm and} ~~
d\mu_{\texttt{inv}}=drd\theta d\phi \sqrt{g_{rr}g_{\theta\theta}
g_{\phi\phi}}.
\]
%
%
The number of modes per invariant volume is divergent at the horizon and
decreases very rapidly as $r$ increases.
Thus we introduce a ultraviolet cutoff at $r_+ + \epsilon$ near the horizon.
The term $\sqrt{g_{rr}g_{\theta\theta}g_{\phi\phi}}$ in the invariant volume
diverge at large distance. Therefore we also introduce an infrared cutoff
at a large distance $L$.

%
The degrees of freedom are mostly concentrated near the horizon, and
the main contribution to the entropy of the system comes from this region.
 This behavior is shown in Fig.~\ref{fig:dof}, where
$r_0$ is a chosen position at which the angular  velocity of ZAMO at
a given $(r, \theta)$ position
$(\Omega_{0} :=\Omega_{0}(r,\theta)=-g_{t\phi}/g_{\phi\phi})$
 differs by $1\% $ from the angular velocity at the horizon $(\Omega_{H}=\Omega_{0}|_{r=r_{+}})$
 to indicate how the degrees of freedom changes as the distance from the horizon changes.

\begin{figure}[pt]
  \includegraphics[width=0.85\textwidth]{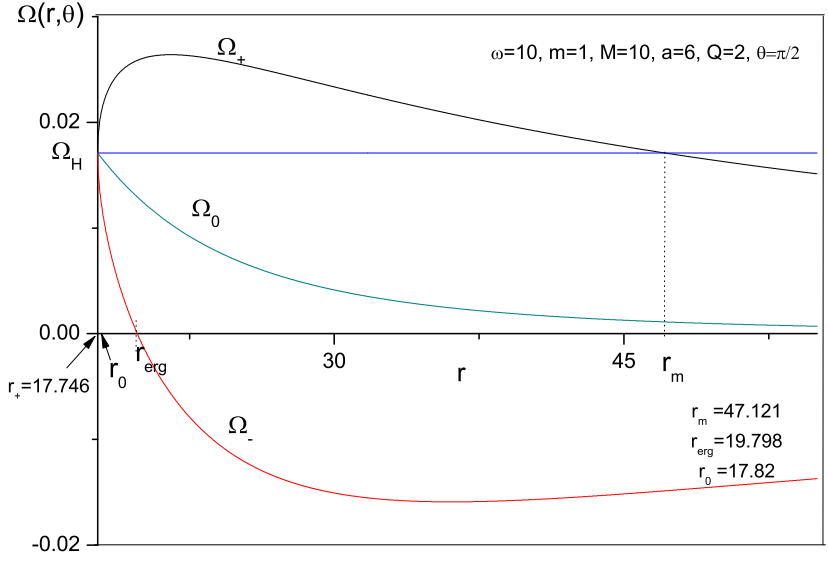}
  \caption{Possible angular velocities of particles are shown at $\theta=\pi/2$.}
    \label{fig:angv}
\end{figure}

Both $\Omega_{\pm}$ converge to  $\Omega_{H}$ near the horizon and vanish at infinity, so the angular velocity of particles near the horizon can be always thought of as $\Omega_{H}$. In particular, the maximum value $\Omega_{+}$ becomes the value $\Omega_{H}$ for a certain radius $r_{m}$. Since the angular velocity
of a particle is less than $\Omega_{H}$ outside $r_{m}$,
thermal equilibrium cannot be achieved for $r > r_m$.
Thus the outer brick-wall should be located inside the radius $r_{m}$,
namely $L<r_{m}$.
This is shown in Fig.~\ref{fig:angv}.

In this paper we assume that thermal equilibrium is maintained in the near horizon region which we are dealing with. Therefore, we can always regard the angular velocity of ZAMO at $(r,\theta)$ inside the brick-wall, $\Omega_{0}(r,\theta)$, to be roughly equal to the horizon angular velocity $\Omega_{H}$, namely $\Omega_{0}(r,\theta)\simeq \Omega_{H}$ near the horizon, i.e., the range of integration giving dominant contributions.

%
%
%
%

We now consider the partition function for the scalar field
following the work of  \cite{hk}:
\begin{eqnarray}
\label{partition}
Z(\beta,\Omega)=\texttt{Tr}e^{-\beta:(\hat{H}-\Omega\hat{J}):},
\end{eqnarray}
where $:\hat{H}:$ and $:\hat{J}:$ are the normal ordered
Hamiltonian and angular momentum operators of the quantized field
and $\beta$ is the inverse temperature $T^{-1}$ with $k=1$.
%
By using the one-particle spectrum, we obtain the free energy from
the partition function as
\begin{eqnarray}
\label{freee}
\beta F=-\ln Z=\sum_{\lambda}\ln\sum_{k}[1-e^{-\beta(\varepsilon_{\lambda}-\Omega j_{\lambda})}]^{k},
\end{eqnarray}
where $\lambda$ denotes the one particle states for the free gas in the system and the occupation
number $k$ takes the values $0,1,2, \cdots$, while $\varepsilon_{\lambda}$ and $j_{\lambda}$ are
expectation values of the normal ordered $\hat{H}$ and $\hat{J}$ in the one particle state $|1_{\lambda}\rangle$.
When we quantize matter fields in a stationary rotating black hole geometry,
we have to take extra care to properly include the contribution from the superradiant modes
which arise due to the presence of the ergoregion.
Previously, the works of \cite{lk:pla,lk:prd,kkps,hkps} failed to include this contribution
properly.
 Following the guideline given in Refs. \cite{mukohyama} and \cite{ford}, we now carry
 the canonical quantization for a scalar field in a rotating black hole system
 from the viewpoint of ROI.
%
%
%
Here, we assume that the set $\varphi_{\omega,m}$ are the mode solutions satisfying the
Klein-Gordon equation (\ref{kle}) and the constraint condition (\ref{reccon}).
We also exclude the mode solutions with the complex frequency
which appeared in the context of vacuum instability related with
the ergoregion \cite{kang,am1,am2,dkm,mdo}.
We define the inner product of the scalar field for its norm
following  \cite{unruh}:
\begin{eqnarray}
\label{definp}
\langle \Phi_{1},\Phi_{2}\rangle =\frac{i}{2}\int_{t=\texttt{const}} \Phi_{1}^{\star}\overleftrightarrow{\partial_{\mu}}\Phi_{2} d\Sigma^{\mu}.
\end{eqnarray}
Using the above definition, we obtain
\begin{eqnarray}
\label{inp}
\langle \varphi_{\omega,m},\varphi_{\omega',m'}\rangle
=\frac{i}{2} \int \varphi_{\omega,m}^{\star}
(\overleftrightarrow{\partial_{t}}+\Omega_{0}\overleftrightarrow{\partial_{\phi}})
\varphi_{\omega',m'}\mathcal{N}^{-1} d\Sigma,
\end{eqnarray}
where $d\Sigma^{\mu}=n^{\mu}d\Sigma$, $n^{\mu}=\mathcal{N}^{-1}(\partial_{t}+\Omega_{0}\partial_{\phi})^{\mu}$,
$\mathcal{N}=(-g^{tt})^{-1/2}$.
Then in the near horizon approximation $\Omega_{0} \simeq \Omega_{H}$,
the norm of a mode solution is given by
\begin{eqnarray}
\label{einp}
\langle \varphi_{\omega,m},\varphi_{\omega,m}\rangle
\simeq \int (\omega-m\Omega_{H})|\varphi_{\omega,m}|^2 \mathcal{N}^{-1} d\Sigma,
\end{eqnarray}
where $\omega \in {\R}$ and $m \in {\Z}$.
In order to make the norm positive $\omega $ and $m $ should satisfy
the following condition:
\begin{eqnarray}
\label{normcd}
 \omega-m\Omega_{H} > 0 .
\end{eqnarray}
%


There are two ways of imposing this condition: One way is to choose the frequency to be positive
following the conventional interpretation as in \cite{kkps,hkps,hk},
and the other way is to leave the frequency to be real as it is given in \cite{kins}.
Now, we explain how the superradiant modes appear in these two approaches.
1) When $ \omega > 0 $, the condition (\ref{normcd}) becomes $ \omega > m\Omega_{H}$ for any $\ m \in {\Z}$.
2) When $ \omega < 0 $, this condition becomes $ 0> \omega > m\Omega_{H}$ for $\ m \in {\Z}_-$ only.
In the conventional approach, the condition 1) corresponds to the non-superradiant modes, and the condition 2)
corresponds to the superradiant modes.
This can be seen by redefining $(\omega,m):=(-\tilde{\omega}, -\tilde{m})$ such that new $(\tilde{\omega}, \tilde{m})$
satisfies the conventional condition for superradiant modes $0<\tilde{\omega} < \tilde{m}\Omega_{H}$ \cite{hk}.
In the second approach \cite{kins}, $ \omega $ is allowed to have any real value and thus no need
 for the above separation.
 However, the superradiant modes should exist in this case too, and they correspond to the negative values of $ \omega $.
%

 In the conventional canonical quantization procedure,
 one adopts the first approach \cite{hk},
 where one sets $(\varepsilon_{\lambda},j_{\lambda})=(\omega,m)$ for
 the non-superradiant(NS) modes and $(\varepsilon_{\lambda},j_{\lambda})=(-\tilde{\omega},-\tilde{m})$
 for the superradiant(SR) modes with $\tilde{\omega}>0, \tilde{m}\in {\Z}_+ $ in Eq.(\ref{freee}).
 Thus, the full set of solutions forms a complete basis with positive frequency
whose norms are  positive definite as follows:
\begin{eqnarray}
\label{cbasis}
\langle \varphi_{\omega,m},\varphi_{\omega',m'} \rangle = \delta({\omega-\omega'})
\delta_{mm'} ~~{\rm with}~ \omega > 0, \ m \in {\Z}, \ \omega > m\Omega_{H} ~~\texttt{for NS}, \nonumber \\
\langle \varphi_{-\tilde{\omega},-\tilde{m}},\varphi_{-\tilde{\omega}',-\tilde{m}'} \rangle = \delta(\tilde{\omega}-\tilde{\omega}')
\delta_{\tilde{m}\tilde{m}'} ~~{\rm with}~ \tilde{\omega}>0, \ \tilde{m}\in {\Z}_+ , \  0<\tilde{\omega} < \tilde{m}\Omega_{H}
 ~~\texttt{for SR}.
\end{eqnarray}
The quantized scalar field can be expanded in terms of normal mode solutions as
\begin{eqnarray}
\label{modeexpo}
\varphi(x)=\sum_{\lambda\in \texttt{NS}}(a_{\omega m}\varphi_{\omega m}+ a_{\omega m}^{\dagger}\varphi_{\omega m}^{\star}
)+\sum_{\lambda\in \texttt{SR}}(a_{-\tilde{\omega},-\tilde{m}}\varphi_{-\tilde{\omega},-\tilde{m}}
+ a_{-\tilde{\omega},-\tilde{m}}^{\dagger}\varphi_{-\tilde{\omega},-\tilde{m}}^{\star}),
\end{eqnarray}
where $\lambda$ denotes the mode set $(\omega,m)$ and $(-\tilde{\omega},-\tilde{m})$ for NS and SR modes, respectively.
 From here on, we remove the tilde from $ \tilde{\omega}$ and $\tilde{m}$ in the SR modes for briefness.

In the second approach \cite{kins},
 all the modes $(\omega,m)$ satisfy $\omega-m\Omega_{H}>0$ giving the positive norm in Eq. (\ref{einp}).
  The frequency $\omega$ can be negative for negative $m$.
Thus the scalar field can be expanded as
\begin{eqnarray}
\label{modeexpt}
\varphi(x)=\sum_{\lambda}(a_{\omega m}\varphi_{\omega m}+ a_{\omega m}^{\dagger}\varphi_{\omega m}^{\star}),
\end{eqnarray}
where the modes $\lambda=(\omega,m)$ satisfy the condition $\omega-m\Omega_{H}>0$ for  $\omega \in {\R}$ and $m \in {\Z}$.
 Given the above analysis, the two approaches should yield the same result.
In this paper, we follow the first approach to evaluate the entropy of the scalar field in the rotating black hole case
which has not been done before and will compare it with the result obtained in \cite{kins}.

In the conventional approach the free energy can be expressed with the sum of the
NS and SR modes,
\begin{eqnarray}
F=F_{\texttt{NS}}+F_{\texttt{SR}},
\end{eqnarray}
where the free energies of the NS and SR modes are given by
\begin{eqnarray}
\label{nsform}
\beta F_{\texttt{NS}}&=&\sum_{\lambda \in \texttt{NS}}\int d\omega ~g(\omega,m)\ln[1-e^{-\beta(\omega-m\Omega_{H})}]
\nonumber \\
&=&\sum_{m}\int_{\omega > m\Omega_{H}} d\omega
~g(\omega,m)\ln[1-e^{-\beta(\omega-m\Omega_{H})}], \\
\label{srform}
\beta F_{\texttt{SR}}&=&\sum_{\lambda \in \texttt{SR}}\int d\omega ~g(\omega,m)\ln[1-e^{\beta(\omega-m\Omega_{H})}]
\nonumber \\
&=&\sum_{m}\int_{\omega < m\Omega_{H}} d\omega
~g(\omega,m)\ln[1-e^{\beta(\omega-m\Omega_{H})}].
\end{eqnarray}
Note that the density functions are given by $g(\omega,m)=\partial n(\omega,m)/\partial \omega$
and $g(\omega,m)=-\partial n(\omega,m)/\partial \omega$ for the NS and SR modes, respectively \cite{kpsy,hk}.

Using the number of modes given by (\ref{moden}) and \eqref{nk},
we now
 calculate the free energy of the total system.
 The free energy for NS modes, Eq. (\ref{nsform}) can be rewritten as
\begin{eqnarray}
\label{expons}
\beta F_{\texttt{NS}} &=& \sum_{m}\int_{\omega > m\Omega_{H}} d\omega
~\frac{\partial}{\partial \omega} \left[\frac{1}{\pi} \int_{r_{+}+\epsilon}^{L} dr k(r;\omega,m)\right]
\ln[1-e^{-\beta(\omega-m\Omega_{H})}]
\nonumber \\
 &=& -\frac{\beta}{\pi} \sum_{m}\int_{r_{+}+\epsilon}^{L} dr\int d\omega
\left[ \frac{k(r;\omega,m)}{e^{\beta(\omega-m\Omega_{H})}-1}\right]
\nonumber \\
&& +\frac{1}{\pi} \sum_{m}\int_{r_{+}+\epsilon}^{L} dr~ k(r;\omega,m)
\ln[1-e^{-\beta(\omega-m\Omega_{H})}|_{\omega_{min}(m)}^{\omega_{max}(m)},
\end{eqnarray}
where we integrated by parts with respect to $\omega$.
For a computational convenience, we divide $F_{NS}$ into two parts
\begin{eqnarray}
\label{calons}
F_{\texttt{NS}}=F_{\texttt{NS}}^{(m >0)}+F_{\texttt{NS}}^{(m <0)},
\end{eqnarray}
which are given by
\begin{eqnarray}
\label{posns}
F_{\texttt{NS}}^{(m >0)}&=&-\frac{1}{\pi} \int\int d\theta d\phi \int_{r_{+}+\epsilon}^{L}dr
\int_{0}^{\infty}d m \int_{m\Omega_{+}}^{\infty}d\omega N^{-2}
~\frac{(\omega-m\Omega_{+})(\omega-m\Omega_{-})}{e^{\beta(\omega-m\Omega_{H})}-1},
\\
\label{negns}
F_{\texttt{NS}}^{(m <0)}&=&-\frac{1}{\pi} \int\int d\theta d\phi\left[\int_{r_{+}+\epsilon}^{r_{erg}}dr
\int_{-\infty}^{0}d m \int_{0}^{\infty}d\omega+ \int_{r_{erg}}^{L}dr \int_{-\infty}^{0}d m \int_{m\Omega_{-}}^{\infty}d\omega \right]
\nonumber \\
 && \times N^{-2}\frac{(\omega-m\Omega_{+})(\omega-m\Omega_{-})}{e^{\beta(\omega-m\Omega_{H})}-1}
\nonumber \\
\!\!&&\!\!-\frac{1}{\pi\beta}\int\int d\theta d\phi \int_{r_{+}+\epsilon}^{r_{erg}}dr
\int_{-\infty}^{0}d m ~N^{-2} m^2~\Omega_{+}\Omega_{-}
\ln(1-e^{\beta m \Omega_{H}}),
\end{eqnarray}
where we regarded the quantum number $m$ as a continuous variable.
Here, we divided the range of $r$ integration into two parts  for the negative $m$ case, since
the minimum angular velocity $\Omega_{-}$ has positive value before $r_{erg}$ and
has negative value beyond $r_{erg}$ as shown in Fig.~\ref{fig:angv}.

On the other hand, the free energy of SR modes, Eq. (\ref{srform}) becomes
\begin{eqnarray}
\label{exposr}
 F_{\texttt{SR}} &=&-\frac{1}{\pi} \int\int d\theta d\phi \int_{r_{+}+\epsilon}^{r_{erg}}dr
\int_{0}^{\infty}d m \int_{0}^{m\Omega_{-}}d\omega
~N^{-2}\frac{(\omega-m\Omega_{+})(\omega-m\Omega_{-})}{e^{-\beta(\omega-m\Omega_{H})}-1}
\nonumber \\
&&+\frac{1}{\pi\beta}\int\int d\theta d\phi \int_{r_{+}+\epsilon}^{r_{erg}}dr
\int_{0}^{\infty}d m~N^{-2}m^2 \Omega_{+}\Omega_{-}\ln(1-e^{-\beta m \Omega_{H}}).
\end{eqnarray}
We note that the second terms in (\ref{negns}) and \eqref{exposr}
 exactly cancel each other in the total free energy as in the BTZ case \cite{hk}.
In evaluating the remaining terms, we adopt the near horizon approximation in which we assume that
the dominant contribution comes only from the radial integration around the near horizon region,
 since in the remaining regions
 the degrees of freedom are negligible as shown in Fig.~\ref{fig:dof}.
Then Eqs. (\ref{posns})-(\ref{exposr}) now yield the following.
\begin{eqnarray}
\label{posnspole}
 F_{\texttt{NS}}^{(m>0)}&=&
 -\frac{\zeta(4)}{\beta^4}
   \Bigg[ \frac{1}{\pi}\int d\theta \frac{(r_{+}^2 +a^2)^4\sin\theta}{(r_{+}-r_{-})^2
 \Sigma_{+}}\left(\frac{1}{\epsilon}
 +F_{2}\ln\left(\frac{r_{+}}{\epsilon}\right) \right)
-\frac{3(r_{+}^2+a^2)^{3}}{a(r_{+}-r_{-})^{3/2}}\frac{1}{\sqrt{\epsilon}} +\mathcal{O}(\sqrt{\epsilon})\Bigg], \nonumber \\
\\
\label{negnspole}
 F_{\texttt{NS}}^{(m<0)}&=&
 -\frac{\zeta(4)}{\beta^4} \Bigg[\frac{2(r_{+}^2+a^2)^{3}}{a(r_{+}-r_{-})^{3/2}}\frac{1}{\sqrt{\epsilon}}
 +\mathcal{O}(\sqrt{\epsilon})\Bigg], \\
\label{srpole}
 F_{\texttt{SR}} &=&
 -\frac{\zeta(4)}{\beta^4}
   \Bigg[ \frac{1}{\pi}\int d\theta \frac{(r_{+}^2 +a^2)^4\sin\theta}{(r_{+}-r_{-})^2
 \Sigma_{+}}\left(\frac{1}{\epsilon}
 +F_{2}\ln\left(\frac{r_{+}}{\epsilon}\right) \right)
+\frac{(r_{+}^2+a^2)^{3}}{a(r_{+}-r_{-})^{3/2}}\frac{1}{\sqrt{\epsilon}} +\mathcal{O}(\sqrt{\epsilon})\Bigg], \nonumber \\
\end{eqnarray}
where
\begin{eqnarray}
\label{F2}
 F_{2}&=& 2\Bigg[\frac{2(\Sigma_{+}r_{-}a^2\sin^2\theta
 +r_{+}[ 2r_{+}^4+r_{+}^2 a^2(5\cos^2\theta-1)
 +a^4 (\cos^4\theta+3\cos^2\theta-2)])}{(r_{+}^2+a^2)^2\Sigma_{+}} \nonumber \\
&&-\frac{1}{r_{+}-r_{-}} \Bigg].
\end{eqnarray}
Using the thermodynamic relation, $S=\beta^{2} \partial F/\partial\beta|_{\beta=\beta_{H}}$,
we obtain the entropy of the system from (\ref{posnspole})-(\ref{srpole}) as
\begin{eqnarray}
\label{entns}
  S_{\texttt{NS}} &=&
 \frac{\zeta(4)}{16\pi^4}\left(\frac{r_{+}-r_{-}}{r_{+}^2+a^2}\right)^3
   \Bigg[ \int d\theta \frac{(r_{+}^2 +a^2)^4\sin\theta}{(r_{+}-r_{-})^2
 \Sigma_{+}}\left(\frac{1}{\epsilon}
 +F_{2}\ln\left(\frac{r_{+}}{\epsilon}\right) \right)
-\frac{\pi(r_{+}^2+a^2)^{3/2}}{a}\frac{1}{\sqrt{\epsilon}}\Bigg],\nonumber \\
\\
\label{entsr}
 S_{\texttt{SR}} &=&
 \frac{\zeta(4)}{16\pi^4}\left(\frac{r_{+}-r_{-}}{r_{+}^2+a^2}\right)^3
   \Bigg[ \int d\theta \frac{(r_{+}^2 +a^2)^4\sin\theta}{(r_{+}-r_{-})^2
 \Sigma_{+}}\left(\frac{1}{\epsilon}
 +F_{2}\ln\left(\frac{r_{+}}{\epsilon}\right) \right)
+\frac{\pi(r_{+}^2+a^2)^{3/2}}{a}\frac{1}{\sqrt{\epsilon}}\Bigg],\nonumber \\
\end{eqnarray}
where we used the relation between the Hawking temperature $\beta_{H}^{-1}$ and the surface
gravity $\kappa_{H}$ of the Kerr-Newman black hole, $\beta_{H}^{-1}
=\kappa_{H}/2\pi=(r_{+}-r_{-})/4\pi (r_{+}^2 +a^2)$.
Note that the $1/\sqrt{\epsilon}$ order terms in (\ref{entns})
and (\ref{entsr}) cancel each other even when $\epsilon$ is $\theta$ dependent.
The total entropy is now given by
\begin{eqnarray}
\label{ent}
 S =S_{\texttt{NS}}+ S_{\texttt{SR}}
 =\frac{\zeta(4)}{8\pi^4}\int d\theta
   \Bigg[ \frac{(r_{+}^2 +a^2)(r_{+}-r_{-})\sin\theta}{
 \Sigma_{+}}\left(\frac{1}{\epsilon}
 +F_{2}\ln\left(\frac{r_{+}}{\epsilon}\right) \right)
+\mathcal{O}(\sqrt{\epsilon})\Bigg],
\end{eqnarray}
where $\Sigma_{+}=r_{+}^2 +a^2 \cos^2 \theta$.
The above obtained total entropy exactly matches with the result of \cite{kins}
except for the following:
 The logarithmically divergent  subleading term is absent  in \cite{kins}.
This is because in \cite{kins} they restricted their concern to black hole singularities
only up to  simple zeros at the horizon.
Thus, if they expanded
the radial integral to the next order,
they would get the same logarithmically divergent  subleading term as in \eqref{ent}.

Finally, by introducing a new cutoff parameter $\bar{\epsilon}$ instead of
the original cutoff parameter $\epsilon$ defined by the following relation,
\begin{eqnarray}
\label{newcut}
\frac{1}{\bar{\epsilon}^2}=\frac{(r_{+}-r_{-})}{2880\pi}
\left(\int_{0}^{\pi} d\theta \frac{\sin\theta}{\Sigma_{+}}\right)\frac{1}{\epsilon}
=\frac{(r_{+}-r_{-})\tan^{-1}(a/r_{+})}{1440\pi~r_{+}a}\frac{1}{\epsilon},
\end{eqnarray}
we can express the total entropy in terms of the surface area of the event horizon,
\begin{eqnarray}
\label{entropy}
S =\Bigg[\frac{A_{H}}{\bar{\epsilon}^2}+
\frac{A_{H}(r_{+}-r_{-})}{720\pi} \left(\int_{0}^{\pi} d\theta \frac{F_{2}(\theta)\sin\theta}{\Sigma_{+}}\right)
\ln\left( \frac{r_{+}}{\bar{\epsilon}}\right)
+\mathcal{O}(\sqrt{\bar{\epsilon}})\Bigg],
\end{eqnarray}
where $A_{H}=4\pi(r_{+}^2+a^2)$ is the surface area of the horizon
and the coefficient of the logarithmically divergent second term
is finite. Choosing the invariant cutoff parameter
$\bar{\epsilon}$ to be twice the planck length
$\bar{\epsilon}=2l_p$ ($l_p^2=G_N$), we retrieve the
Bekenstein-Hawking relation in the leading order.

\section{Entropy from the viewpoint of ZAMO near horizon}
In this section we calculate the entropy from the viewpoint of ZAMO near the horizon.
We first consider the near horizon line element of the Kerr-Newman black hole in the coordinates rotating  with angular velocity of the black hole.
Now we rewrite the Kerr-Newman metric (\ref{metric}) as
\begin{eqnarray}
\label{remetric}
ds^2 =\frac{1}{g^{tt}}dt^2+g_{rr}dr^2 +g_{\theta\theta}d\theta^2
+g_{\phi\phi}(d\phi+ \Omega_{0} dt)^2,
\end{eqnarray}
where $\Omega_{0}(r,\theta)=-g_{t\phi}/g_{\phi\phi}$.
Using the coordinate transformation $\varphi = \phi-\Omega_{H}t$
we change the metric (\ref{remetric}) into the coordinates rotating with
the angular velocity of the horizon, that is given by
\begin{eqnarray}
\label{reremetric}
ds^2 = \frac{1}{g^{tt}}dt^2 +g_{rr}dr^2 +g_{\theta\theta}d\theta^2
+g_{\phi\phi}(d\varphi-(\Omega_{0}-\Omega_{H})dt)^2.
\end{eqnarray}
The above metric is diagonal only in the vicinity of the horizon due to the coordinate dependence of the $\Omega_{0}(r,\theta)$
which becomes $\Omega_{H}$ at the horizon.

On the one hand, as we saw in the previous section the dominant contribution to
the entropy comes only from the region near the horizon. Since we only consider
the region near horizon we now rewrite the metric (\ref{reremetric})
in the vicinity of the horizon,
\begin{eqnarray}
\label{nhmetric}
ds^2 \approx \frac{1}{g^{tt}}dt^2 +g_{rr}dr^2 +g_{\theta\theta}d\theta^2
+g_{\phi\phi}d\varphi^2.
\end{eqnarray}

In the WKB approximation with $\Psi(x)=e^{-i E t+im\varphi+iK(r,\theta)}$,
the field equation (\ref{kle}) with (\ref{nhmetric}) gives the constraint condition
\begin{eqnarray}
\label{cconzamo}
\frac{p_{r}^2}{g_{rr}}+\frac{p_{\theta}^2}{g_{\theta\theta}}
=(-g^{tt})(E^2-m^2 \tilde{\Omega}^2),
\end{eqnarray}
where $\tilde{\Omega}=(-g^{tt}g_{\phi\phi})^{-1/2}$.
Here we note that the above constraint condition for the momenta is the same as in
(\ref{ccon}) obtained from the ROI's viewpoint if we set $E:=\omega-m \Omega_{H}$.

On the other hand, a ZAMO near horizon needs to measure the physical quantities locally
in evaluating the entropy of the system. This observer measures local inverse
temperature as $\beta=\beta_H \sqrt{-g'_{tt}}$ and energy as $\varepsilon=E/\sqrt{-g'_{tt}}$,
where $\beta_H$ is the inverse Hawking temperature of the black hole
measured by the ROI and
$g'_{tt}:=g_{tt}+2g_{t\phi}\Omega_{H}+g_{\phi\phi}\Omega_{H}^2=1/g^{tt}$ \cite{mi98}.
Here, the combined product  $\beta \varepsilon$ from the ZAMO's viewpoint
remains the same as $\beta_H E$ from the ROI's viewpoint as it was discussed in \cite{mi98}.
 To assume thermal equilibrium near the horizon, we
  regard the local inverse temperature $\beta$ to be approximately constant
  near the horizon as we did in the ROI case.
Now we change the temperature and energy from the ROI viewpoint to the local temperature and energy measured by the ZAMO near the horizon, and then
the constraint condition (\ref{cconzamo}) can be rewritten as
\begin{eqnarray}
\label{ccontzamo}
\frac{p_{r}^2}{g_{rr}}+\frac{p_{\theta}^2}{g_{\theta\theta}}
=\left(\varepsilon^2-\frac{m^2}{g_{\phi\phi}}\right).
\end{eqnarray}
The number of modes $n(\varepsilon,m)$ with energy less than $\varepsilon$ and with a fixed $m$ can be calculated by
integrating over the phase space.
In the present case, the number of modes is given by
\begin{eqnarray}
\label{modenzamo}
\pi n(\varepsilon,m)=
\int_{r_{+}+\epsilon}^{L} dr k(r; \varepsilon,m),
\end{eqnarray}
where the radial wave number $k(r; \varepsilon,m)$ can be evaluated
as in the previous section and is given by as follows
with the restriction
that the right hand side of Eq.\eqref{ccontzamo} be positive:
\begin{eqnarray}
\label{k}
k(r; \varepsilon,m)=\frac{1}{8\pi^2}\int\int d\theta d\varphi
dp_{\theta}~dp_{r}= \frac{1}{8\pi} \int \int d\theta d\phi~ (g_{rr}g_{\theta\theta})^{1/2}
 \left(\varepsilon^2-\frac{m^2}{g_{\phi\phi}}\right) .
\end{eqnarray}

The free energy is then given by
\begin{eqnarray}
\label{exponszamo}
\beta F &=& \sum_{m}\int d\varepsilon
~\frac{\partial}{\partial \varepsilon} \left[\frac{1}{\pi} \int_{r_{+}+\epsilon}^{L} dr k(r;\varepsilon,m)\right]
\ln[1-e^{-\beta\varepsilon}]
\nonumber \\
 &=& -\frac{\beta}{\pi} \sum_{m}\int_{r_{+}+\epsilon}^{L} dr\int d\varepsilon
\left[ \frac{k(r;\varepsilon,m)}{e^{\beta\varepsilon}-1}\right]
\nonumber \\
&& +\frac{1}{\pi} \sum_{m}\int_{r_{+}+\epsilon}^{L} dr~ k(r;\varepsilon,m)\ln[1-e^{-\beta\varepsilon}]|_{\varepsilon_{min}(m)}^{\varepsilon_{max}(m)}.
\end{eqnarray}
Note that there is no contribution from superradiant modes
 since there is no rotation of the frame from ZAMO's viewpoint.
For convenience, we also divide the free energy into $m>0$ and $m<0$ parts
due to the restriction that the right hand side of Eq.\eqref{cconzamo} be positive,
\begin{eqnarray}
F=F^{(m>0)}+F^{(m<0)},
\end{eqnarray}
where  the two parts are given by
\begin{eqnarray}
\label{posnszamo}
F^{(m >0)}&=&-\frac{1}{\pi} \int\int d\theta d\varphi \int_{r_{+}+\epsilon}^{L}dr
\int_{0}^{\infty}d m \int_{m/\sqrt{g_{\phi\phi}}}^{\infty}d\varepsilon~
~\frac{k(r;\varepsilon,m)}{e^{\beta\varepsilon}-1},
\\
\label{negnszamo}
F^{(m <0)}&=&-\frac{1}{\pi} \int\int d\theta d\varphi \int_{r_{+}+\epsilon}^{L}dr
\int_{-\infty}^{0}d m \int_{-m/\sqrt{g_{\phi\phi}}}^{\infty}d\varepsilon~
~\frac{k(r;\varepsilon,m)}{e^{\beta\varepsilon}-1}.
\end{eqnarray}
Now, the total free energy can be written as
\begin{eqnarray}
\label{freeszamo}
F &=&-\frac{2}{\pi} \int\int d\theta d\varphi \int_{r_{+}+\epsilon}^{L}dr
\int_{0}^{\infty}d m \int_{m/\sqrt{g_{\phi\phi}}}^{\infty}d\varepsilon~
~\frac{k(r;\varepsilon,m)}{e^{\beta\varepsilon}-1},
\nonumber \\
&=& -\frac{1}{4\pi^2} \int_{r_{+}+\epsilon}^{L}dr \int\int d\theta d\varphi
(g_{rr}g_{\theta\theta})^{1/2}
\int_{0}^{\infty}d m G_{m}|_{m/\sqrt{g_{\phi\phi}}}^{\infty},
\end{eqnarray}
where
\begin{eqnarray}
\label{gm}
G_{m}(\varepsilon)=\int d\varepsilon \frac{\left(\varepsilon^2-\frac{m^2}{g_{\phi\phi}}\right)}{e^{\beta\varepsilon}-1}.
\end{eqnarray}
The integration  \eqref{gm} can be done straightforwardly, and we get
\begin{eqnarray}
\int_{0}^{\infty} dm~G_{m}|_{m/\sqrt{g_{\phi\phi}}}^{\infty}=
\frac{2\zeta(4)\Gamma(4)}{3\beta^4}
(g_{\phi\phi})^{1/2}.
\end{eqnarray}
Thus, the total free energy is given by
\begin{eqnarray}
\label{freezamo}
F &=&-\frac{\zeta(4)}{\pi^2} \int_{r_{+}+\epsilon}^{L}dr \int\int d\theta d\varphi
\frac{(g_{rr}g_{\theta\theta}g_{\phi\phi})^{1/2}} {\beta^4}.
\end{eqnarray}
Using the thermodynamic relation $S=\beta^2\frac{\partial F}{\partial \beta}
|_{\beta=\tilde{\beta}}$, the entropy of the system is given by
\begin{eqnarray}
\label{entzamo}
 S &=& \frac{4\zeta(4)}{\pi^2 }\int_{r_{+}+\epsilon}^{L} dr d\theta d\phi \frac{\sqrt{g_{rr}g_{\theta\theta}g_{\phi\phi}}}
 { \tilde{\beta}^3} \nonumber \\
 &=& \frac{4\zeta(4)}{\pi^2\beta_{H}^3 }\int_{r_{+}+\epsilon}^{L} dr d\theta d\phi
 (-g^{tt})^{3/2}(g_{rr}g_{\theta\theta}g_{\phi\phi})^{1/2},
 \end{eqnarray}
where we used the relation $\tilde{\beta}=\beta_{H}/\sqrt{-g^{tt}}$.
Evaluating the entropy with the near horizon approximation
as we did in the ROI case,
 \eqref{entzamo} becomes
\begin{eqnarray}
\label{fentzamo}
 S =\frac{\zeta(4)}{8\pi^4}\int d\theta
   \Bigg[ \frac{(r_{+}^2 +a^2)(r_{+}-r_{-})\sin\theta}{
 \Sigma_{+}}\left(\frac{1}{\epsilon}
 +F_{2}\ln\left(\frac{r_{+}}{\epsilon}\right) \right)
+\mathcal{O}(\sqrt{\epsilon})\Bigg],
\end{eqnarray}
where $F_{2}$ is given by  (\ref{F2}) in the previous section.
This result exactly coincides with the total entropy (\ref{ent})
obtained from the ROI's viewpoint. Again introducing the invariant
cutoff parameter $\bar{\epsilon}$ defined by \eqref{newcut} in the
previous section and setting $\bar{\epsilon}=2l_p$, we also obtain
the Bekenstein-Hawking relation in the leading order.

\section{Discussion}

In this paper, using the brick-wall model with a scalar field in
the canonical quantization approach we investigated the entropy of
the Kerr-Newman black hole from two different viewpoints, a rest
observer at infinity (ROI) and a zero angular momentum
observer(ZAMO) near horizon. The results from the two viewpoints
coincide exactly. This is what we expected since the total entropy
of the system must be a same physical quantity regardless of an
observer.

 As it is well known,
the superradiant modes occur due to the presence of the ergoregion
in a rotating black hole system.
Incorporating the superradiant modes and evaluating
the entropy of rotating black hole in the brick-wall model
was a bit complicated and thus raised some confusion. This issue was not
settled down until the work of Ho and Kang \cite{hk}.
They evaluated the entropy of three dimensional rotating BTZ black hole
correctly first time in the brick-wall model.
Rather recently in the work of \cite{kins} the result of \cite{hk} in the
three dimensional rotating BTZ black hole case was reproduced
using a rather different approach in the brick-wall model,
and using the same method they also obtained the entropy of four dimensional
Kerr black hole.
However, the approach of \cite{kins} was not quite conventional in such a way that
negative frequencies are also allowed as superradiant modes.
In the conventional approach of \cite{hk} only positive frequencies are allowed,
and evaluating the contribution of the superradiant modes in terms of positive
frequencies was rather tricky.
In the conventional treatment for the superradiant modes,
the frequencies satisfy the condition $0<\omega< m\Omega_H$, and
in the calculation of the free energy  $(\omega, ~ m)$
should be regarded as  $(-\omega, ~ -m)$ as first pointed out in \cite{hk}.
Before the work of \cite{hk} people simply used $(\omega, ~ m)$ instead of $(-\omega, ~ -m)$
in the evaluation of the free energy,  thus obtained a divergent contribution from the superradiant modes \cite{kkps,hkps}.
While in the approach of \cite{kins} by allowing negative frequencies
 the superradiant modes were not separated from the non-superradiant modes in evaluating the free energy,
 thus the calculation becomes  simplified quite a bit.
Still it was not certain whether the approach of \cite{hk} would yield the same result
 as in  \cite{kins} in the four dimensional Kerr black hole case, since the relation
 between the two approaches was not understood clearly so far.
In this paper, we explain how the approaches of \cite{hk} and \cite{kins}
can be understood on the same footing and show that the approach of \cite{hk}
applied to the four dimensional Kerr black hole case actually yields the same
result as in \cite{kins}.

%
It is also well known that
the superradiant modes do not occur
in a rotation free co-moving system. Thus if we evaluate
the entropy in this rotation free co-moving system (ZAMO)
we are free from considering the troublesome superradiant modes.
Therefore we naturally expect that the evaluation of rotating black hole entropy would be quite simpler
from ZAMO's viewpoint.
The entropy calculation from ZAMO's viewpoint has not been performed so far,
thus leaving the question that whether the entropy from ZAMO's viewpoint
actually agrees with the known result from ROI's viewpoint unanswered.
We confirm this in the latter part of this work.


We expect that if we apply this equivalence between the ROI's and ZAMO's
in evaluating the entropy in the case of Kerr-Newman-de Sitter black hole
we would get the result in a quite simpler fashion without complication
of the superradiant modes \cite{ely}.



\section*{Acknowledgments}
The authors thank KIAS for hospitality during the time that this work was done.
This work was supported by the Korea Research Foundation grant funded by
the Korea Government(MOEHRD) KRF-2006-312-C00498(E. C.-Y. and D. L.), and
 by the Korea Science and
Engineering Foundation grant funded by the Korea
government(MEST) through the Center for Quantum Spacetime(CQUeST)
 of Sogang University with grant number
R11-2005-021(M. Y.).


\end{document}